\documentclass[runningheads]{llncs}
\usepackage[T1]{fontenc}
\usepackage{graphicx,verbatim}
\usepackage{longtable}
\newlength\savewidth\newcommand\shline{\noalign{\global\savewidth\arrayrulewidth
  \global\arrayrulewidth 1.25pt}\hline\noalign{\global\arrayrulewidth\savewidth}}
\usepackage{xspace}
\usepackage{threeparttable}
\usepackage{multirow}
\newcommand{\methodname}{\texttt{AR-Seg}\xspace}
\usepackage{amsmath,amsfonts}
\usepackage{amssymb,amsopn}
\usepackage{algorithm}
\usepackage{algpseudocode}
\usepackage{bm}
\usepackage{bbm}

\newcommand{\std}[1]{}

\newcommand{\eat}[1]{}

\newcommand{\tabref}[1]{Table~\ref{#1}}
\newcommand{\figref}[1]{Fig.~\ref{#1}}

\newcommand{\tabincell}[2]{\begin{tabular}{@{}#1@{}}#2\end{tabular}}

\newcommand{\loss}[1]{\mathcal{L}_\text{#1}}

\newcommand{\best}[1]{\textbf{#1}}
\newcommand{\suboptimal}[1]{\underline{#1}}

\newcommand{\img}[1]{\boldsymbol{#1}}

\usepackage[colorlinks=true,
    linkcolor=blue,
urlcolor=blue,citecolor=blue]{hyperref}
\begin{document}
\title{Autoregressive Medical Image Segmentation \\ via Next-Scale  Mask Prediction}
\titlerunning{Autoregressive Medical Image Segmentation}
\author{Tao Chen, Chenhui Wang, Zhihao Chen, Hongming Shan}
\institute{Institute of Science and Technology for Brain-inspired Intelligence\\
Fudan University}
\maketitle
\begin{abstract}
While deep learning has significantly advanced medical image segmentation, most existing methods still struggle with handling complex anatomical regions. 
Cascaded or deep supervision-based approaches attempt to address this challenge through multi-scale feature learning but fail to establish sufficient inter-scale dependencies, as each scale relies solely on the features of the immediate predecessor.
Towards this end, we propose the  \underline{A}uto\underline{R}egressive \underline{Seg}mentation framework via next-scale mask prediction, termed \methodname, which progressively predicts the next-scale mask by explicitly modeling dependencies across all previous scales within a unified architecture.
\methodname introduces three innovations: 
(1) a multi-scale mask autoencoder that quantizes the mask into multi-scale token maps to capture hierarchical anatomical structures, 
(2) a next-scale autoregressive mechanism that progressively predicts next-scale masks to enable sufficient inter-scale dependencies, and 
(3) a consensus-aggregation strategy that combines multiple sampled results to generate a more accurate mask, further improving segmentation robustness.
Extensive experimental results on two benchmark datasets with different modalities demonstrate that \methodname outperforms state-of-the-art methods while explicitly visualizing the intermediate coarse-to-fine segmentation process.


\end{abstract}
\section{Introduction}
Medical image segmentation is crucial for clinical diagnosis and treatment planning~\cite{haque2020deep}.
Despite advancements in deep learning, most existing methods~\cite{chen2021transunet,chen2023berdiff,isensee2021nnu} struggle with the variability of complex anatomical structures---differing in size, shape, and appearance---making accurate segmentation particularly challenging in intricate and ambiguous regions.
To address these challenges, cascaded and deep supervision-based approaches~\cite{aghapanah2025mecardnet,zhang2018deep} have focused on multi-scale feature integration or loss constraints to extract more robust representations. However, these methods still struggle to establish sufficient inter-scale dependencies, as each scale relies solely on the features of the immediate predecessor.

Drawing from the success of autoregressive models via next-token prediction in natural language processing \cite{radford2018improving}, we recognize their potential to address the inherent variability in discrete segmentation tasks. To this end, we propose a novel \underline{A}uto\underline{R}egressive \underline{Seg}mentation framework via next-scale mask prediction, termed \methodname, which progressively predicts the next-scale mask by explicitly modeling dependencies across all previous scales within a unified architecture. This holistic modeling of multi-scale dependencies leads to a more coherent and accurate segmentation, especially in regions with varying anatomical features, thus improving both segmentation robustness and clinical usability.
To the best of our knowledge, \methodname is the first time to utilize next-scale prediction~\cite{tian2024visual} for the discrete segmentation task.

Our contributions are summarized as follows. \textbf{(i)} We propose \methodname, an autoregressive framework via next-scale mask prediction for medical image segmentation, explicitly modeling dependencies across all previous scales within a unified architecture. \textbf{(ii)} We introduce a multi-scale mask autoencoder that quantizes the mask into multi-scale token maps, capturing hierarchical anatomical structures. \textbf{(iii)} We present a next-scale autoregressive mechanism that progressively predicts next-scale mask, enabling sufficient dependencies across scales. \textbf{(iv)} We introduce a consensus-aggregation strategy that combines multiple sampled results into a final mask, further improving segmentation robustness. \textbf{(v)} Extensive experiments on two benchmark datasets with different modalities demonstrate that \methodname outperforms state-of-the-art (SOTA) methods while explicitly visualizing the intermediate coarse-to-fine segmentation process.
\section{Methodology}
\begin{figure}[!t]
\centering
\includegraphics[width=1\linewidth]{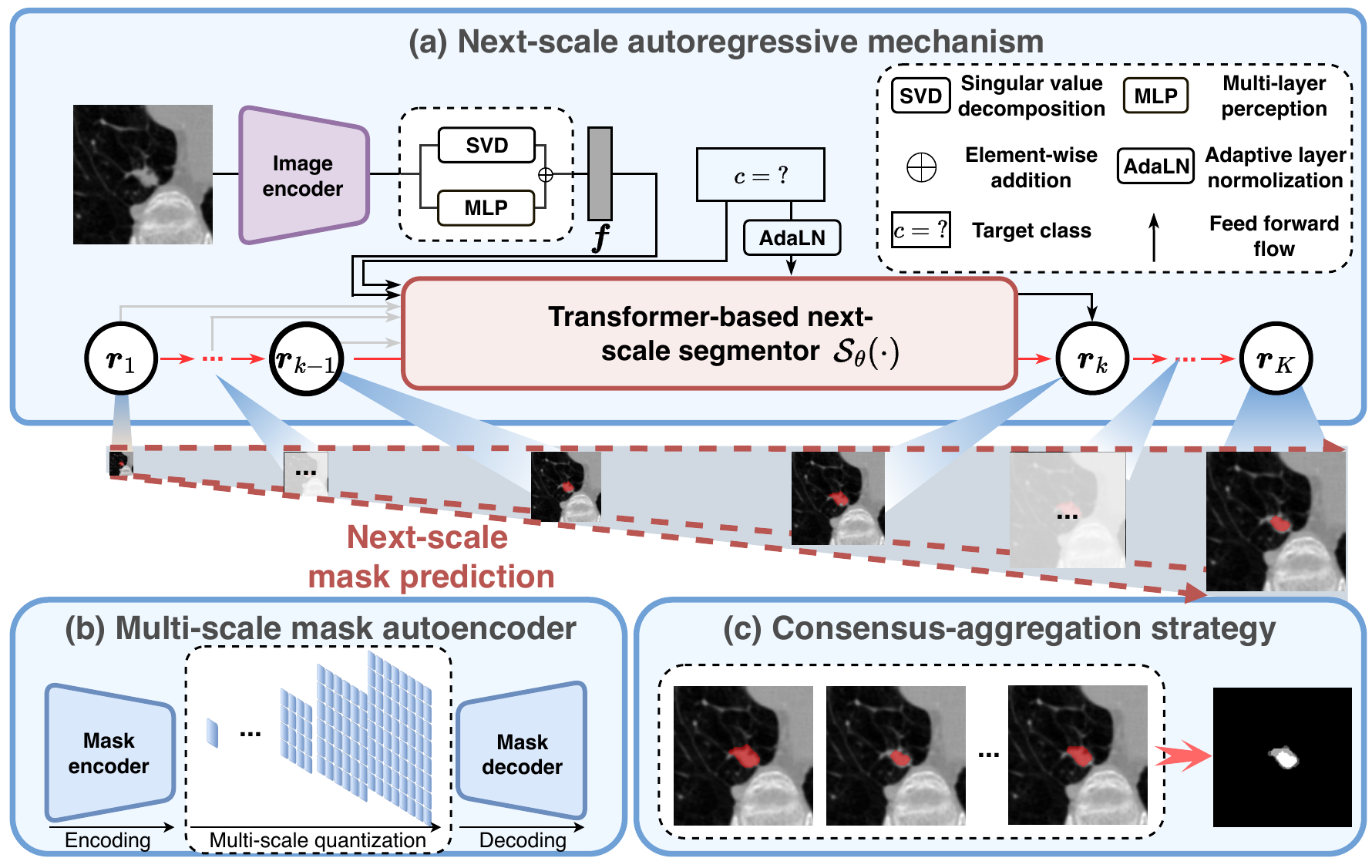}
  \caption{Illustration of the proposed \methodname.}
  \label{fig:overviews}
\end{figure}

Let $\img{x}$ and $\img{y}$ denote the medical image and its corresponding mask with $C$ target classes, respectively. \methodname predicts the mask $\hat{\img{y}}$ in a next-scale autoregressive manner, as shown in \figref{fig:overviews}.
Specifically, we first quantize $\img{y}$ into $K$ multi-scale token maps via a multi-scale mask autoencoder, then apply a next-scale autoregressive mechanism to predict the next-scale mask progressively. Finally, a consensus-aggregation strategy produces the final mask by combining multiple sampled results. We then detail each component along with the training and inference processes.
\subsection{Multi-Scale Mask Autoencoder}
\label{sec:mask}
To effectively capture hierarchical anatomical structures while preserving critical discriminative information across scales, we propose a multi-scale mask autoencoder to progressively quantize the mask to token maps instead of aggressively downsampling it. 
The autoencoder consists of a mask encoder $\mathcal{E}_{\text{mask}}(\cdot)$, a mask decoder $\mathcal{D}_{\text{mask}}(\cdot)$, and a token quantizer $\mathcal{Q}(\cdot)$.

\noindent\textbf{Mask quantized process.}\quad We first encode the mask $\img{y}$ into a feature map $\img{m}=\mathcal{E}_{\text{mask}}(\img{y})$. 
Then, we dynamically update the feature map $\img{m}$ in a residual manner~\cite{lee2022autoregressive,tian2024visual} to enhance the token differences across $K$ scales from coarse to fine, and progressively quantize $\img{m}$ into $K$ multi-scale token maps, $\mathcal{R} = (\img{r}_1, \img{r}_2, \dots, \img{r}_K)$.
Specifically, the $k$-scale quantized process maps each element of the mask residual feature map $\img{m}$ to the nearest code index $\img{r}_k^{(i,j)}$ from a learnable codebook $\img{Z}$ of size $V$:
\begin{align}
    {\img{r}_k}^{(i, j)} = \mathcal{Q}(\img{m}^{(i,j)})= {\arg \min }_{v \in[V]}\|\operatorname{lookup}(\img{Z}, v)-{\img{m}}^{(i, j)}\|_{2},
\end{align}
where $\operatorname{lookup}(\img{Z}, v)$ means taking the $v$-th vector in codebook $\img{Z}$. The detailed quantized process is shown in Alg.~\ref{alg:encoding}.

\noindent\textbf{Mask dequantized process.}\quad Based on the multi-scale token maps $\mathcal{R}$, the dequantized process first progressively dequantizes the token map to the corresponding mask residual feature map $\hat{\img{m}}$ and then uses the mask decoder $\mathcal{D}_{\text{mask}}(\cdot)$ to reconstruct the mask $\hat{\img{y}}$, as shown in Alg.~\ref{alg:decoding}.

\begin{figure}[t]
\centering
\begin{minipage}{0.49\linewidth}
\begin{algorithm}[H]
\footnotesize
\caption{Quantized process}\label{alg:encoding}
\begin{algorithmic}[1]
\State \textbf{Inputs:} mask $\img{y}$, scales $K$, resolutions $(h_k, w_k)^{K}_{k=1}$;
\State $\img{m}=\mathcal{E}_{\text{mask}}(\img{y})$, $\mathcal{R}=[]$;
\For{$k = 1$ to $K$}
\State $\img{r}_k = \mathcal{Q}(\text{interpolate}(\img{m}, h_k, w_k))$;
\State $\mathcal{R} = \text{queue\_push}(\mathcal{R}, \img{r}_k)$;
\State $\img{z}_k = \text{lookup}(\img{Z}, \img{r}_k)$;
\State $\img{z}_k = \text{interpolate}(\img{z}_k, h_K, w_K)$;
\State $\img{m} = \img{m} - \phi_{k}(\img{z}_k)$;
\EndFor
\State \Return multi-scale token maps $\mathcal{R}$
\end{algorithmic}
\end{algorithm}
\end{minipage}
\hfil
\begin{minipage}{0.49\linewidth}
\begin{algorithm}[H]
\footnotesize
\caption{Dequantized process}\label{alg:decoding}
\begin{algorithmic}[1]
\State \textbf{Inputs:} multi-scale token maps $\mathcal{R}$, scales $K$, resolutions $(h_k, w_k)^{K}_{k=1}$;
\State $\hat{\img{m}} = \img{0}$;
\For{$k = 1$ to $K$}
\State $\img{r}_k = \text{queue\_pop}(\mathcal{R})$;
\State $\img{z}_k = \text{lookup}(\img{Z}, \img{r}_k)$;
\State $\img{z}_k = \text{interpolate}(\img{z}_k, h_K, w_K)$;
\State $\hat{\img{m}} = \hat{\img{m}} + \phi_{k}(\img{z}_k)$;
\EndFor
\State $\hat{\img{y}} = \mathcal{D}_{\text{mask}}(\hat{\img{m}})$;
\State \Return reconstructed mask $\hat{\img{y}}$
\end{algorithmic}
\end{algorithm}
\end{minipage}
\end{figure}

\subsection{Next-Scale Autoregressive Mechanism}
\label{sec:ar}
To explicitly model inter-scale dependencies in a unified architecture, we propose a next-scale autoregressive mechanism based on multi-scale token maps. This mechanism employs a transformer-based~\cite{vaswani2017attention} next-scale segmentor to reformulate the medical image segmentation task as a conditional autoregressive process.

Specifically, we first use the pretrained MedSAM~\cite{MedSAM} as the image encoder backbone, $\mathcal{E}_{\text{image}}(\cdot)$, followed by a parameter-efficient adapter to extract the corresponding image features.
Instead of using a non-linear multi-layer perceptron (MLP) directly, our adapter enhances the feature extraction through linear singular value decomposition (SVD)~\cite{klema1980singular} to preserve more global image information, producing the final image embedding $\img{f}$ as:
\begin{align}
    \img{f} = \text{MLP}\left(\mathcal{E}_{\text{image}}(\img{x})\right) + \text{SVD}\left(\mathcal{E}_{\text{image}}(\img{x})\right).
\end{align}

Then, for each scale $k$, the next-scale segmentor $\mathcal{S}_\theta(\cdot)$ with parameters $\theta$ takes the target class $c$, the image embedding $\img{f}$, and all previously predicted mask token maps as input to predict the $k$-scale token map. This process defines the autoregressive likelihood of the segmentation as follows:
\begin{align}
    p(\img{r}_{1}, \img{r}_{2}, \ldots, \img{r}_{K})={\textstyle\prod_{k=1}^{K}}p_{\theta}(\img{r}_{k}| \img{r}_{1}, \img{r}_{2}, \ldots, \img{r}_{k-1}, c, \img{f}).
\end{align}
Note that the target class is also passed through adaptive layer normalization (AdaLN)~\cite{karras2019style} to condition the segmentor. We highlight that this next-scale autoregressive mechanism can explicitly visualize the intermediate coarse-to-fine segmentation process, helping clinicians improve AI confidence.

\subsection{Consensus-Aggregation Strategy}
\label{sec:con}
To further improve segmentation performance and resolve ambiguity in challenging regions, we introduce a consensus-aggregation strategy. 
Specifically, we first sample $N$ multi-scale token maps, $\mathcal{R}^n|_{n=1}^N$. For each element $\img{r}_{k}^{n}$ in $\mathcal{R}^{n}$, we sample it from a multinomial distribution $\mathcal{M}$, with probabilities generated by the next-scale segmentor:
\begin{align}
\img{r}_{k}^{n} \sim \mathcal{M}\Big(\img{r}_{k}^{n}; \mathcal{S}_{\theta}(\img{r}_{1}, \img{r}_{2}, \ldots, \img{r}_{k-1}, c, \img{f})\Big).
\end{align}
We then dequantize these $N$ multi-scale token maps $\mathcal{R}^{n}|_{n=1}^{N}$ into the corresponding masks $\hat{\img{y}}^n|_{n=1}^{N}$ through Alg.~\ref{alg:decoding}. Finally, these multiple sampled results are aggregated using a mean operation: $\hat{\img{y}} = \frac{1}{N}\sum_{n=1}^{N} \hat{\img{y}}^{n} $, which enhances the model's ability to handle ambiguous regions. 

\subsection{Detailed Procedure}
\label{sec:train}
Here, we provide the detailed training and inference procedures of our \methodname. 

The training of \methodname consists of two stages: multi-scale mask learning and next-scale autoregressive learning.
In the first stage, the multi-scale mask autoencoder is trained by minimizing a compound loss $\loss{}$, which combines a quantization constraint and a segmentation objective:
\begin{align}
    \loss{} =\underbrace{\|\img{m}\!-\!\text{sg}({\hat{\img{m}}})\|_{2} \!+\! \beta\|\text{sg}(\img{m})\!-\!\hat{\img{m}}\|_{2}}_{\text{quantization constraint}}+\underbrace{\lambda_{\mathrm{Dice}} \mathcal{L}_{\mathrm{Dice}}(\img{y},  \hat{\img{y}})\!+\!\lambda_{\mathrm{BCE}} \mathcal{L}_{\mathrm{BCE}}(\img{y}, \hat{\img{y}})}_{\text{segmentation objective}}
\end{align}
where $\beta$ is the weight for the commitment loss~\cite{esser2021taming}, $\text{sg}(\cdot)$ denotes the stop-gradient operation, while $\mathcal{L}_{\mathrm{Dice}}$ and $\mathcal{L}_{\mathrm{BCE}}$ are the Dice~\cite{milletari2016v} and binary cross-entropy losses, with $\lambda_{\mathrm{Dice}}$ and $\lambda_{\mathrm{BCE}}$ as their respective weights. In the second stage, we train the next-scale segmentor optimized with cross-entropy loss~\cite{radford2018improving} with the frozen multi-scale mask autoencoder. A block-wise causal attention mask is applied during training to ensure causality.

During inference, the next-scale segmentor iteratively generates token maps based on the class and image embedding until the multi-scale token maps are formed, which are then dequantized using Alg.~\ref{alg:decoding} and passed through the consensus-aggregation strategy to produce the final segmentation mask.

\section{Experiment}
\subsection{Experimental Setup}
\noindent\textbf{Dataset and preprocessing.}\quad The experiments are conducted on two datasets: LIDC-IDRI~\cite{clark2013cancer} and BRATS~2021~\cite{Baid2021TheRB}. The LIDC-IDRI dataset comprises 1,018 lung CT scans, each annotated by four radiologists. These CT scans are preprocessed using a standard pipeline and follow a train-validation-test split consistent with prior studies~\cite{chen2023berdiff,kohl2018probabilistic}. Note that we randomly select one of four annotated masks for each image during training. The BRATS 2021 dataset provides four MRI modalities (T1, T2, FLAIR, and T1CE) per patient, which are concatenated along the channel axis to form a combined input for our experiments. The preprocessing and train-test split follow previous studies~\cite{chen2024hidiff}. Segmentation performance is evaluated on three tumor types: necrotic tumor core (NT), peritumoral edema (ED), and enhancing tumor (ET).

\noindent\textbf{Implementation details.}\quad All experiments are implemented by PyTorch~\cite{paszke2019pytorch} and trained on NVIDIA V100 GPUs. The multi-scale mask autoencoder is trained for 100 epochs with a batch size of 128. While the next-scale segmentor is trained for 300 epochs with a batch size of 32. Both networks employ the AdamW optimizer~\cite{loshchilov2017decoupled} with an initial learning rate of $1 \times 10^{-4}$ and a cosine annealing schedule. The loss weights $\beta$, $\lambda_{\mathrm{Dice}}$, and $\lambda_{\mathrm{BCE}}$, are empirically set to $0.25$, $1$, and $1$, respectively. The number of scale $K$ is set to 8.  For performance evaluation, we employ four metrics: Generalized Energy Distance (GED)~\cite{kohl2018probabilistic}, Hungarian-Matched Intersection over Union (HM-IoU)~\cite{kohl2019hierarchical}, Soft-Dice~\cite{Ji2021LearningCM}, and Dice coefficient. GED is computed using varying numbers of segmentation samples (1, 4, 8, and 16), while HM-IoU and Soft-Dice are calculated using 16 samples.

\begin{figure}[!t]
\centering
\begin{minipage}[t]{0.6\linewidth}
\makeatletter\def\@captype{table}
\caption{Results on LIDC-IDRI.}
\renewcommand\arraystretch{0.9}
\label{tab:quantitative_CT}
\begin{threeparttable}
\begin{tabular*}{1\linewidth}{@{\extracolsep{\fill}}lccc}
\shline
\textbf{Methods} &\tabincell{c}{\textbf{GED}$\downarrow$\\\textbf{16}}
& \tabincell{c}{\textbf{HM-IoU}$\uparrow$\\\textbf{16}} & \tabincell{c}{\textbf{Soft-Dice}$\uparrow$\\\textbf{16}}\\
\cline{1-4}
Prob.U-net~\cite{kohl2018probabilistic} &  0.320\std{$\pm$0.03}   & 0.500\std{$\pm$0.03} & -\\
Hprob.U-net~\cite{kohl2019hierarchical} &  0.270\std{$\pm$0.01}   & 0.530\std{$\pm$0.01} & 0.624\std{$\pm$0.01}\\
D-Personal~\cite{wu2024diversified}&0.346\std{$\pm$0.01}&0.555\std{$\pm$0.02}&0.604\std{$\pm$0.01}\\
CAR~\cite{kassapis2021calibrated}       &  0.264\std{$\pm$0.00}    &0.592\std{$\pm$0.01} & 0.633\std{$\pm$0.00}\\
MR-Net~\cite{Ji2021LearningCM} & 0.658 & 0.447 & 0.616\\
PADL~\cite{liao2024modeling} & 0.544 & 0.462 & 0.595\\
MedSegDiff~\cite{wu2022medsegdiff}& 0.420\std{$\pm$0.03}  &  0.413\std{$\pm$0.03} & 0.453\std{$\pm$0.02} \\
BerDiff~\cite{chen2023berdiff} & \suboptimal{0.238\std{$\pm$0.01}}  &  \suboptimal{0.596\std{$\pm$0.00}} & \suboptimal{0.644\std{$\pm$0.00}}\\
PixelSeg~\cite{zhang2022pixelseg}       &  0.260\std{$\pm$0.00} &0.587\std{$\pm$0.01} & - \\
\methodname (\textbf{ours})&\best{0.232\std{$\pm$0.00}}&\best{0.616\std{$\pm$0.01}}&\best{0.658\std{$\pm$0.01}}\\
\shline
\end{tabular*}
\end{threeparttable}
\end{minipage}
\begin{minipage}[t]{0.3\linewidth}
\makeatletter\def\@captype{table}
\caption{Results on BRATS~2021.}
\label{tab:quantitative_MRI}
\begin{threeparttable}
\begin{tabular*}{1\linewidth}{@{\extracolsep{\fill}}lc}
\shline
\textbf{Methods}&\textbf{Dice}$\uparrow$\\
\cline{1-2}
U-net~\cite{ronneberger2015u}& 75.48 \\
nnU-net~\cite{isensee2021nnu} & 84.57 \\
TransU-net~\cite{chen2021transunet} & 83.50 \\
SwinU-net~\cite{cao2022swin}&83.94\\
MERIT~\cite{rahman2023multi}& 83.02 \\
MedSegDiff~\cite{amit2021segdiff} & 82.81 \\
BerDiff~\cite{chen2023berdiff}&85.42\\
HiDiff~\cite{chen2024hidiff} & \suboptimal{85.80} \\
\methodname (\textbf{ours})&\best{86.97}\\
\shline
\end{tabular*}
\end{threeparttable}
\end{minipage}
\end{figure}

\begin{figure}[!t]
\centering
  \includegraphics[width=1\linewidth]{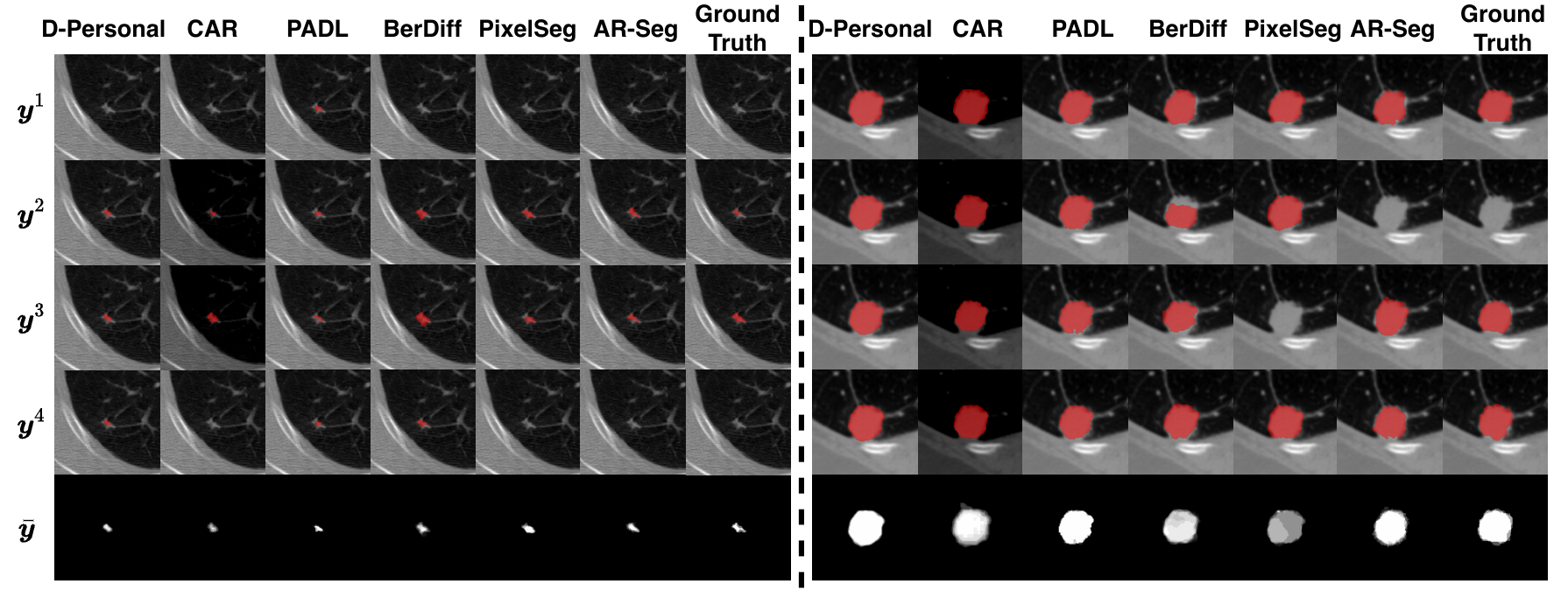}
  \caption{\textbf{Qualitative results of two lung nodules from LIDC-IDRI.}
   $\img{y}^{i}$ and $\bar{\img{y}}$ refer to the $i$-th segmentation masks and the final consensus-aggregated masks, respectively.
}
  \label{fig:visualization_LIDC}
\end{figure}

\subsection{Experimental Results}
\noindent\textbf{Comparison to SOTA methods.}\quad We evaluate our \methodname against other SOTA methods on two datasets: LIDC-IDRI, to assess performance under inter-radiologist variability, and BRATS 2021, to evaluate multi-class segmentation of small lesions and ambiguous boundaries, using different performance metrics.

\noindent\textit{(i) Results on LIDC-IDRI.}\quad We compare \methodname with conditional variational autoencoder-based methods like Prob. U-net~\cite{kohl2018probabilistic}, Hprob. U-net~\cite{kohl2018probabilistic}, and D-Personal~\cite{wu2024diversified}, generative adversarial network-based  CAR~\cite{kassapis2021calibrated}, multi-annotator variability modeling techniques like MR-Net~\cite{Ji2021LearningCM} and PADL~\cite{liao2024modeling}, diffusion-based models like MedSegDiff~\cite{wu2022medsegdiff} and BerDiff~\cite{chen2023berdiff}, and pixel-wise autoregressive PixelSeg~\cite{zhang2022pixelseg}.
From the quantitative results in \tabref{tab:quantitative_CT}, 
we draw three key observations as follows. (1) While many SOTA methods leverage U-shape cascaded architectures to capture multi-scale features, their lack of inter-scale dependencies hampers the segmentation. (2) Diffusion based methods, despite their explicit mask distribution modeling, still overlook multi-scale feature integration, resulting in lower accuracy compared to \methodname. (3) By incorporating next-scale prediction and explicit inter-scale dependency modeling, \methodname outperforms pixel-wise PixelSeg. Qualitative results are shown in \figref{fig:visualization_LIDC}. Compared to other models, \methodname accurately captures segmentation patterns. For the left example, \methodname generates masks that align more closely with the ground truth, particularly in capturing subtle prominences at the lower right of the nodules.

\begin{figure}[!t]
\centering
  \includegraphics[width=1\linewidth]{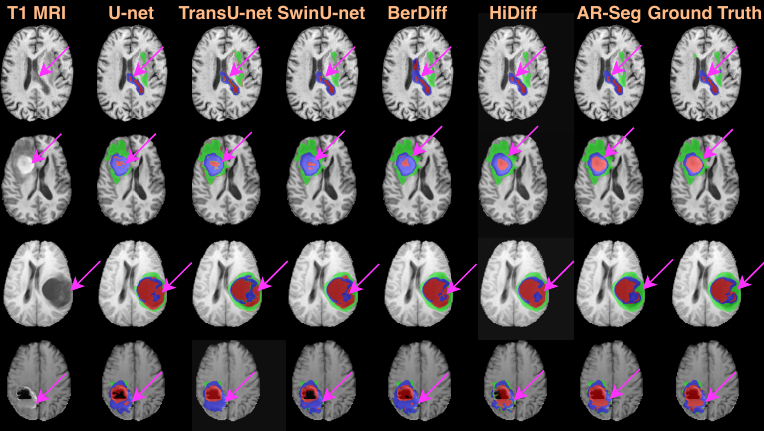}
  \caption{\textbf{Qualitative results of four MRI images from BRATS~2021.}  Only T1-weighted images are shown for convenience.}
  \label{fig:visualization_BRATS}
\end{figure}

\begin{table}[!t]
\centering
\renewcommand\arraystretch{0.9}
\caption{Ablation results of multi-scale mask autoencoder.}\label{tab:ablation_progressive}
\begin{threeparttable}
\begin{tabular*}{1\linewidth}{@{\extracolsep{\fill}}ccccccccc}
\shline
&\multicolumn{4}{c}{\textbf{Single-scale}}&\multicolumn{4}{c}{\textbf{Multi-scale}}\\
\cline{2-5} \cline{6-9}
\textbf{Class}&Nodules&NT&ED&ET&Nodules&NT&ED&ET\\
\textbf{Dice}$\uparrow$&0.997&0.991&0.976&0.987&\best{0.999}&\best{0.996}&\best{0.988}&\best{0.993}\\
\shline
\end{tabular*}
\end{threeparttable}
\end{table}

\noindent\textit{(ii) Results on BRATS~2021.}\quad
We present quantitative and qualitative results in \tabref{tab:quantitative_MRI} and \figref{fig:visualization_BRATS}, respectively, comparing \methodname with convolution-based methods (U-net~\cite{ronneberger2015u} and nnU-net~\cite{isensee2021nnu}), transformer-based approaches (TransU-net~\cite{chen2021transunet}, SwinU-net~\cite{cao2022swin}, and MERIT~\cite{rahman2023multi}), diffusion-based models (MedSegDiff~\cite{wu2022medsegdiff} and BerDiff~\cite{chen2023berdiff}) and hybrid diffusion method (HiDiff~\cite{chen2024hidiff}). 
First, while transformer-based methods enhance image feature extraction, they fail to address the inherent limitation of U-shaped architectures in modeling inter-scale dependencies, resulting in inferior performance. Besides, although hybrid diffusion methods refine masks using diffusion processes, their single-scale operation limits their effectiveness, making them less accurate than \methodname. From \figref{fig:visualization_BRATS}, we observe that \methodname achieves improved accuracy in ambiguous regions. For instance, in the 3rd row, \methodname accurately segments a challenging tumor region.

\noindent\textbf{Ablation study.}\quad 
We first conduct ablation experiments to evaluate the effectiveness of the multi-scale mask autoencoder in \tabref{tab:ablation_progressive}. We introduce a single-scale baseline, where the mask features are quantized into a single scale. The results demonstrate that the multi-scale mask autoencoder achieves higher Dice coefficients across all classes, owing to its multi-scale design, which enables learning a more robust representation through inter-scale mutual reinforcement.

We then conduct ablation studies on the next-scale autoregressive mechanism using LIDC-IDRI in \tabref{tab:ablation_component}. We compare against a baseline that uses a single-scale mask autoencoder and generates the mask via next-token prediction. When shifting the next-token to next-scale objective, our \methodname largely outperforms this baseline, demonstrating the benefits of the next-scale progression. Further gains are observed by replacing the MLP adapter with our SVD-based adapter, highlighting its contribution to more robust feature extraction.

\begin{table}[!t]
\centering
\renewcommand\arraystretch{0.9}
\caption{Ablation results of next-scale autoregressive mechanism on LIDC-IDRI.}\label{tab:ablation_component}
\begin{threeparttable}
\begin{tabular*}{1\linewidth}{@{\extracolsep{\fill}}lccccc}
\shline
\multirow{2}{*}{\textbf{Framework}}&\multicolumn{4}{c}{\textbf{GED}$\downarrow$}&\textbf{HM-IoU}$\uparrow$\\
&\textbf{16}&\textbf{8}&\textbf{4}&\textbf{1}&\textbf{16}\\
\shline
Baseline&0.275&0.310&0.383&0.814&0.557\\
\quad+ Next-scale autoregressive&0.239&0.277&0.348&0.745&0.595\\
\quad\quad + SVD-based adapter&\best{0.232}&\best{0.267}&\best{0.332}&\best{0.736}&\best{0.616}\\
\shline
\end{tabular*}
\end{threeparttable}
\end{table}

\begin{figure}[t]
\centering
\includegraphics[width=1\linewidth]{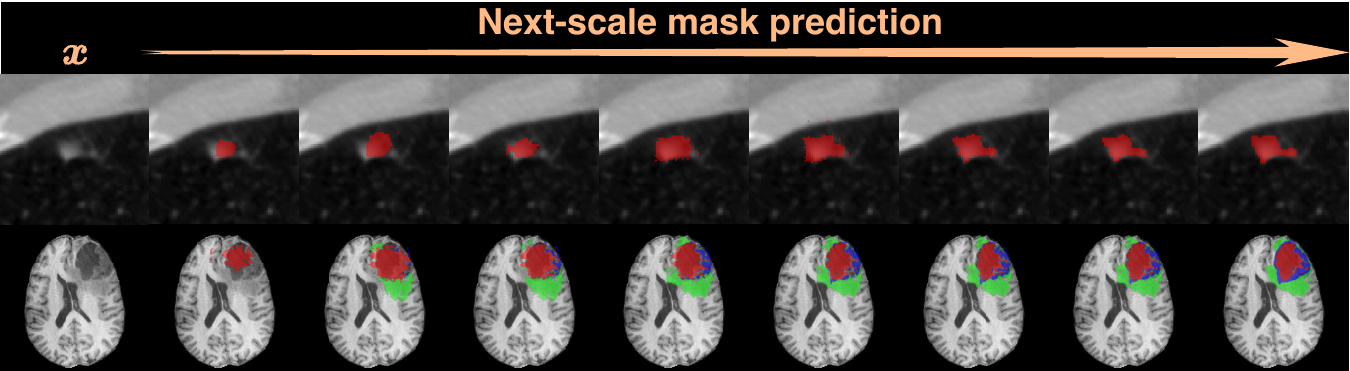}
  \caption{Explicit segmentation process of \methodname.}
  \label{fig:interpretable}
\end{figure}

\noindent\textbf{Explicit segmentation process.}\quad We highlight the explicit segmentation advantage of \methodname, demonstrated in the qualitative results in \figref{fig:interpretable}, where the process starts with an initial coarse mask and iteratively refines the boundaries. For example, \methodname enhances the contextual relationships between tumor subtypes in brain tumors as the scale increases.
\section{Conclusion}
This paper introduces the autoregressive segmentation framework via next-scale mask prediction, a novel approach for medical image segmentation that explicitly models inter-scale dependencies. By incorporating the multi-scale mask autoencoder, next-scale autoregressive mechanism, and the consensus-aggregation strategy, \methodname not only achieves accurate segmentation but also provides visualization of the intermediate coarse-to-fine segmentation process. Experiments on the LIDC-IDRI and BRATS~2021 show that \methodname outperforms SOTA methods. Future work will focus on integrating \methodname into clinical practice by exploring clinician-in-the-loop strategies to leverage its progressive capabilities.


\begin{thebibliography}{10}
\providecommand{\url}[1]{\texttt{#1}}
\providecommand{\urlprefix}{URL }
\providecommand{\doi}[1]{https://doi.org/#1}

\bibitem{aghapanah2025mecardnet}
Aghapanah, H., Rasti, R., Tabesh, F., Pouraliakbar, H., Sanei, H., Kermani, S.: {MECardNet}: A novel multi-scale convolutional ensemble model with adaptive deep supervision for precise cardiac {MRI} segmentation. Biomed. Signal Process. Control.  \textbf{100},  106919 (2025)

\bibitem{amit2021segdiff}
Amit, T., Shaharbany, T., Nachmani, E., Wolf, L.: {SegDiff}: Image segmentation with diffusion probabilistic models. arXiv:2112.00390  (2021)

\bibitem{Baid2021TheRB}
Baid, U., Ghodasara, S., Mohan, S., Bilello, M., Calabrese, E., Colak, E., Farahani, K., Kalpathy-Cramer, J., Kitamura, F.C., Pati, S., et~al.: The {RSNA-ASNR-MICCAI BraTS} 2021 benchmark on brain tumor segmentation and radiogenomic classification. arXiv:2107.02314  (2021)

\bibitem{cao2022swin}
Cao, H., et~al.: {Swin-Unet}: {Unet-like} pure transformer for medical image segmentation. In: ECCV. pp. 205--218 (2022)

\bibitem{chen2021transunet}
Chen, J., Lu, Y., Yu, Q., Luo, X., Adeli, E., Wang, Y., Lu, L., Yuille, A.L., Zhou, Y.: {TransUNet}: Transformers make strong encoders for medical image segmentation. arXiv:2102.04306  (2021)

\bibitem{chen2024hidiff}
Chen, T., Wang, C., Chen, Z., Lei, Y., Shan, H.: {HiDiff}: {H}ybrid diffusion framework for medical image segmentation. IEEE Trans. Med. Imaging  \textbf{43}(10),  3570--3583 (2024)

\bibitem{chen2023berdiff}
Chen, T., Wang, C., Shan, H.: {BerDiff}: Conditional {Bernoulli} diffusion model for medical image segmentation. In: MICCAI (2023)

\bibitem{clark2013cancer}
Clark, K., Vendt, B., Smith, K., Freymann, J., Kirby, J., Koppel, P., Moore, S., Phillips, S., Maffitt, D., Pringle, M., et~al.: The cancer imaging archive ({TCIA}): maintaining and operating a public information repository. J. Digit. Imaging  \textbf{26}(6),  1045--1057 (2013)

\bibitem{esser2021taming}
Esser, P., Rombach, R., Ommer, B.: Taming transformers for high-resolution image synthesis. In: CVPR. pp. 12873--12883 (2021)

\bibitem{haque2020deep}
Haque, I.R.I., Neubert, J.: Deep learning approaches to biomedical image segmentation. Informatics Med. Unlock.  \textbf{18},  100297 (2020)

\bibitem{isensee2021nnu}
Isensee, F., Jaeger, P.F., Kohl, S.A., Petersen, J., Maier-Hein, K.H.: {nnU-Net}: a self-configuring method for deep learning-based biomedical image segmentation. Nat. Methods  \textbf{18}(2),  203--211 (2021)

\bibitem{Ji2021LearningCM}
Ji, W., Yu, S., Wu, J., Ma, K., Bian, C., Bi, Q., Li, J., Liu, H., Cheng, L., Zheng, Y.: Learning calibrated medical image segmentation via multi-rater agreement modeling. In: CVPR. pp. 12336--12346 (2021)

\bibitem{karras2019style}
Karras, T., Laine, S., Aila, T.: A style-based generator architecture for generative adversarial networks. In: CVPR. pp. 4401--4410 (2019)

\bibitem{kassapis2021calibrated}
Kassapis, E., Dikov, G., Gupta, D.K., Nugteren, C.: Calibrated adversarial refinement for stochastic semantic segmentation. In: ICCV. pp. 7037--7047 (2020)

\bibitem{klema1980singular}
Klema, V., Laub, A.: The singular value decomposition: Its computation and some applications. IEEE Trans. Autom. Control  \textbf{25}(2),  164--176 (1980)

\bibitem{kohl2018probabilistic}
Kohl, S., Romera-Paredes, B., Meyer, C., De~Fauw, J., Ledsam, J.R., Maier-Hein, K., Eslami, S., Jimenez~Rezende, D., Ronneberger, O.: A probabilistic {U-Net} for segmentation of ambiguous images. In: NIPS (2018)

\bibitem{kohl2019hierarchical}
Kohl, S.A., Romera-Paredes, B., Maier-Hein, K.H., Rezende, D.J., Eslami, S., Kohli, P., Zisserman, A., Ronneberger, O.: A hierarchical probabilistic {U-Net} for modeling multi-scale ambiguities. In: medical workshop of NIPS (2019)

\bibitem{lee2022autoregressive}
Lee, D., Kim, C., Kim, S., Cho, M., Han, W.S.: Autoregressive image generation using residual quantization. In: CVPR. pp. 11523--11532 (2022)

\bibitem{liao2024modeling}
Liao, Z., Hu, S., Xie, Y., Xia, Y.: Modeling annotator preference and stochastic annotation error for medical image segmentation. Med. Image Anal.  \textbf{92},  103028 (2024)

\bibitem{loshchilov2017decoupled}
Loshchilov, I., Hutter, F.: Decoupled weight decay regularization. In: ICLR (2017)

\bibitem{MedSAM}
Ma, J., He, Y., Li, F., Han, L., You, C., Wang, B.: Segment anything in medical images. Nat. Commun.  \textbf{15}, ~654 (2024)

\bibitem{milletari2016v}
Milletari, F., Navab, N., Ahmadi, S.A.: V-net: Fully convolutional neural networks for volumetric medical image segmentation. In: 3DV. pp. 565--571 (2016)

\bibitem{rahman2023multi}
Mostafijur~Rahman, M., Marculescu, R.: Multi-scale hierarchical vision transformer with cascaded attention decoding for medical image segmentation. In: MIDL (2023)

\bibitem{paszke2019pytorch}
Paszke, A., Gross, S., Massa, F., Lerer, A., Bradbury, J., et~al.: {PyTorch}: An imperative style, high-performance deep learning library. In: NIPS (2019)

\bibitem{radford2018improving}
Radford, A., Narasimhan, K.: Improving language understanding by generative pre-training (2018), \url{https://api.semanticscholar.org/CorpusID:49313245}

\bibitem{ronneberger2015u}
Ronneberger, O., Fischer, P., Brox, T.: {U-Net}: Convolutional networks for biomedical image segmentation. In: MICCAI. pp. 234--241 (2015)

\bibitem{tian2024visual}
Tian, K., Jiang, Y., Yuan, Z., Peng, B., Wang, L.: Visual autoregressive modeling: Scalable image generation via next-scale prediction. In: NIPS (2024)

\bibitem{vaswani2017attention}
Vaswani, A., Shazeer, N., Parmar, N., Uszkoreit, J., Jones, L., Gomez, A.N., Kaiser, {\L}., Polosukhin, I.: Attention is all you need. In: NIPS (2017)

\bibitem{wu2022medsegdiff}
Wu, J., FU, R., Fang, H., Zhang, Y., Yang, Y., Xiong, H., Liu, H., Xu, Y.: {MedSegDiff}: Medical image segmentation with diffusion probabilistic model. In: MIDL. pp. 1623--1639 (2024)

\bibitem{wu2024diversified}
Wu, Y., Luo, X., Xu, Z., Guo, X., Ju, L., Ge, Z., Liao, W., Cai, J.: Diversified and personalized multi-rater medical image segmentation. In: CVPR. pp. 11470--11479 (2024)

\bibitem{zhang2022pixelseg}
Zhang, W., Zhang, X., Huang, S., Lu, Y., Wang, K.: {PixelSeg}: Pixel-by-pixel stochastic semantic segmentation for ambiguous medical images. In: ACM MM. pp. 4742--4750 (2022)

\bibitem{zhang2018deep}
Zhang, Y., Chung, A.C.: Deep supervision with additional labels for retinal vessel segmentation task. In: MICCAI. pp. 83--91 (2018)

\end{thebibliography}
\end{document}